\documentclass[12pt]{iopart}
\usepackage{epsfig}
\usepackage{graphicx}

\begin{document}
\newcommand{\avg}[1]{\langle{#1}\rangle}
\newcommand{\Avg}[1]{\lamgle{#1} \rangle}
\def\be{\begin{equation}}
\def\ee{\end{equation}}
\def\bea{\begin{eqnarray}}
\def\eea{\end{eqnarray}}
\title[Percolation transition and distribution of connected components]{Percolation transition and distribution of connected components in generalized random network ensembles}
\author{Serena Bradde$^1$ and Ginestra Bianconi$^2$ }
\address{$^1$ International School for Advanced Studies, via Beirut
  2/4, 34014, Trieste, Italy and INFN, Via Valerio 2, Trieste, Italy}
\address{$^2$ The Abdus Salam International Center for Theoretical Physics, Strada Costiera 11, 34014, Trieste, Italy}
\ead{{\mailto{bradde@sissa.it}, \mailto{gbiancon@ictp.it}}}
\begin{abstract}
In this work, we study the percolation transition and large deviation properties of
generalized canonical network ensembles. This new type of random networks
might have a very rich complex structure, including high heterogeneous degree
sequences, non-trivial community structure or specific spatial dependence of
the link probability for networks embedded in a metric space. We find the
cluster distribution of the networks in these ensembles by mapping the problem
to a fully connected Potts model with heterogeneous couplings. We show that
the nature of the Potts model phase transition, linked to the birth of a giant
component, has a crossover from second to first order when the number of
critical colors $q_c = 2$ in all the networks under study. These results shed light
on the properties of dynamical processes defined on these network ensembles.

\end{abstract}

\vspace{2pc}
\maketitle
\section{Introduction}

Recently the study of critical phenomena in complex networks has attracted 
a great deal of interest \cite{Dorogovtsev}.
. One of the main critical phenomena occurring in networks is the percolation
transition which is a continuous structural phase transition that can be characterized by
critical indices as a statistical mechanics second-order phase transition. This phase transition
determines the robustness properties of complex networks 
\cite{MR,Attack,Cohen1,Cohen2} and the critical temperature  of the Ising
\cite{Ising1,Ising2,Ising3} and XY models \cite{Isaac,Coolen} on
complex networks.
 Moreover, the onset of a percolating cluster determines a transition 
in between a phase in which small loops are
suppressed and a phase in which the expectation value of small loops is 
positive in the limit of large network sizes  \cite{Noh}.

The percolation phase transition in Erd\"os and Renyi networks is a
classic subject of graph theory \cite{Bollobas}. For this network
ensembles the large
deviation of the number of connected components (or clusters) 
has been characterized \cite{Monasson} by a mapping of the problem to a
fully connected Potts model \cite{Fortuin}.

In uncorrelated complex networks, characterized by a non-Poisson degree
distribution, the percolation transition depends on the
second moment of the degree distribution \cite{Cohen1, Cohen2} and
can show non trivial critical exponents \cite{Dorogovtsev}.

This phase transition has been also studied in directed networks
\cite{Boguna} and in networks  with degree-degree correlations
\cite{Doro2}.

In this paper we study the percolation properties and the large
deviation of the cluster distribution of the recently proposed generalized
canonical random network ensembles \cite{entropy1,entropy2} with non trivial degree distribution   and an  additional  community
structure or spatial structure. These networks ensembles can be cast
in the wide category of  Configuration or  ``hidden variable'' models
extensively study in the recent literature \cite{MR,hv1,hv2,hv3,hv4,hv5} .
The percolation properties and the large deviations of the cluster
distribution in these ensembles are studied in this paper  by  mapping
the problem   to a fully connected
Potts model with heterogeneous couplings. We find  results in
agreement with reference \cite{Lee} where the Potts model formulation was first used for  the study of the
percolation properties of complex networks with heterogeneous degrees.
In particular our  framework generalize the results of \cite{Lee} and can be
applied in network ensembles with very diverse structure, not only
network ensembles with
heterogeneous degree distribution, but also network ensembles with an
additional non-trivial community or  spatial structure.

The paper is organized as follows. In section 2 we introduce the
generalized canonical random ensembles. In section 3  we introduce the
generating functions for the cluster distribution and we characterize
its large deviations. In section 4 we relate the problem of finding
the cluster distribution in generalized canonical model, and their
percolation transition,   to the study
of a fully connected  Potts model with heterogeneous couplings.
In section 5 we solve the fully connected Potts model with
heterogeneous couplings and we find the percolation threshold and
critical exponent $\hat{\beta}$ for the generalized canonical network
ensembles. In section 6 we find the cluster distribution in the generalized
canonical network ensembles. In section 7 we compare our theoretical
predictions with simulation results. Finally in section 8 we give the conclusions.

\section{Random network ensembles }

In this section we introduce the generalized  random ensembles described
in \cite{entropy1,entropy2}. The generalized random ensembles are an
extension of the known $G(N,M)$ and $G(N,p)$ random network
ensembles and are related to Configuration and "hidden variable''
ensembles \cite{MR,hv1,hv2,hv3,hv4,hv5}. 

\subsection{The $G(N,M)$ and $G(N,p)$ random network ensembles}

The mathematical literature has widely studied the properties of the
$G(N,M)$ and $G(N,p)$ random network ensembles.
\begin{itemize}
\item
A random network in the ensemble  $G(N,M)$ is a network having $N$ nodes 
and $M$  undirected links. 
If we indicate with $a_{ij}$ the adjacency matrix of the network (with
$a_{ij}=1$ if there is a link between node $i$ and $j$ and $a_{ij}=0$ 
otherwise), the probability that a network ${\cal G}$, associated to the 
adjacency matrix ${\bf a}$, belongs to the $G(N,M)$ ensemble is given by
\be
P({\cal G})=\frac{1}{Z}\delta(M,\sum_{i<j}a_{ij})
\ee
with 
\bea
Z=\left(\begin{array}{c} \frac{N(N-1)}{2}\\
M\end{array}\right).
\eea
and with the $\delta(\cdot)$ indicating the Kronecker delta.
The probability of  each link in this ensemble of networks is given 
by $p= M/(N(N-1)/2)$.
\item A network in the $G(N,p)$ ensemble is a network in which 
each possible pair of links is present with probability $p$. Therefore the 
probability of  a specific network ${\cal G}$ in this ensemble is equal to
\begin{equation}\label{prob}
P_C({\cal G})= \prod_{i<j}p^{a_{ij}}(1-p)^{1-a_{ij}}
\end{equation}
where $a_{ij}$ is the adjacency matrix.
In the $G(N,p)$ ensemble the total number of links $M$ is not fixed
 but is Poisson distributed with mean $\avg{M}_{P_C(G)}=pN(N-1)/2$.
\end{itemize}
The $G(M,N)$ and the $G(N,p)$ ensemble with $p=M/(N(N-1)/2)$
are linked by a Legendre transform, and, in the asymptotic limit of
$N\rightarrow\infty$, they share the same statistical properties.

\subsection{Generalized random network ensembles}

Recently a statistical mechanics approach has been proposed 
\cite{entropy1,entropy2} that is able
to generalize the  random networks ensembles to network ensembles with much more
complex structure including networks with highly heterogeneous degree
sequences and non trivial community structure or spatial dependence of
the link probability.
The statistical mechanics approach is able to describe both
"microcanonical" network ensembles (that satisfy hard structural
constraints and generalize the $G(N,M)$ random ensembles) and
"canonical" network ensembles (that satisfy the structural
constraints when their properties are averaged over the whole ensemble
and generalize the $G(N,p)$ ensemble).
\begin{itemize}
\item The "microcanonical" networks have to satisfy a
series of hard constraints $\vec{F}({\cal G})=\vec{C}$ and the probability of these
networks are given by
\be
P_{MC}({\cal G})=\frac{1}{Z}\delta(\vec{F}({\cal G})-\vec{C})
\ee
with $Z$ indicating the cardinality of the ensemble.
The probability  of each link $p_{ij}$ is computed introducing some
Lagrange multipliers \cite{entropy1,entropy2}.
\item
The "canonical" conjugated ensemble can be built starting from the
probability of the links $p_{ij}$ in the ``microcanonical'' one. We
assign to each network ${\cal G}$ the probability
\begin{equation}
P_C({\cal G})=\prod_{i<j}p_{ij}^{a_{ij}}(1-p_{ij})^{1-a_{ij}}\,.
\label{canonical}
\end{equation}
which generalizes (\ref{prob}) to heterogeneous networks.
In the "canonical" ensembles the structural constraints are satisfied
on average
\be
\overline{\vec{F}({\cal G})}=\vec{C}.
\ee
Here and in the following we always indicate by $\overline{\cdots}$
the average over the ensemble probability $P_C({\cal G})$ given by
$(\ref{canonical})$ and with $\avg{\cdots}$ the average over
all the nodes $i=1,\ldots, N$.
\end{itemize}
In this paper we focus on  generalized "canonical" networks.
Each node $i$  in this ensemble is characterized by two discrete \emph{hidden
variables} $\theta_i$ and $\alpha_i$. We consider in this paper the link
probability given by 
\be
p_{ij}=\frac{\theta_i \theta_jW(\alpha_i,\alpha_j)}{1+\theta_i\theta_j
  W(\alpha_i,\alpha_j)}
\label{ens}
\ee
and is fully specified once the  function $W(\alpha,\alpha')$ is given.
The link probability (\ref{ens}) corresponds to maximally entropic
ensembles with given degree structural constraints
\cite{entropy1,entropy2}.

In the ensembles described by (\ref{ens}), the  degree of each
node $k_i$ is a Poisson variable \cite{hv5} with average
\be
\overline{k}_i=\sum_{j\neq i} p_{ij}.
\ee

In the following we specifically comment on some relevant limiting
cases for the general distribution $(\ref{ens})$.
\begin{itemize}
\item
{\it The $G(N,p)$ ensemble}\\
If the values of the \emph{hidden variables} $\theta$'s are equal,
i.e. $\theta_i=\theta\;$ $\forall i$ and $W(\alpha,\alpha')=\delta_{\alpha,\alpha'}$, the probability of a link  is given by 
\be
p_{ij}=p=\theta^{2}/(1+\theta^2)
\ee
The degree of each node is a  Poisson variable with equal average
$\overline{k}=pN$. Performing also the average $\avg{\cdot}$
over all the nodes of the network we get
\be
p=\frac{\avg{\overline{k}}}{N}
\ee
We recover therefore the Erd\"os and Renyi ensemble $G(N,p)$ by
taking 
\be
\theta=\sqrt{\frac{\avg{\overline{k}}/N}{1-\avg{\overline{k}}/N}}\simeq
\sqrt{\avg{\overline{k}}/N}
\ee
where the last  expression is valid for sparse networks with $\avg{\overline{k}}$ finite.
\item
{\it The Configuration model}\\
If the linking probability $p_{ij}$ of equation $(\ref{ens})$ depends only on
$\theta_i$ and $\theta_j$, (i.e. $W(\alpha,\alpha')=1$ $\alpha,
\alpha'$), then 
\be
p_{ij}=\frac{\theta_i\theta_j}{1+\theta_i\theta_j}.
\ee
This ensemble is the canonical version of the Configuration model each node
$i$ having a degree $k_i$  distributed according to a Poisson
variable with average 
\be
\overline{k}_i=\sum_j \frac{\theta_i \theta_j}{1+\theta_i \theta_j}.
\label{cc}
\ee
This ensemble has in general non-trivial degree degree correlations
that disappears for  $\max_i(\theta_i)\ll 1$.
In this last case, $\max_i(\theta_i)\ll 1$ the linking probability
$p_{ij}$  defined in equation $(\ref{cc})$ can be approximated as
\be
p_{ij}=\theta_i\theta_j.
\label{ucorr}
\ee
Therefore in this limit the  networks of the ensemble  are uncorrelated and there is a simple relation
between the hidden variables $\theta_i $ and the average degree
$\overline{k}_i$ of the node $i$, i.e.
\be
\theta_i=\frac{\overline{k}_i}{\sqrt{\avg{\overline{k}}N}}.
\label{thetak}
\ee
Finally we observe that if we use $(\ref{thetak})$ the linking
probability $p_{ij}$ can be expressed in the well known expression for
uncorrelated networks
\be
p_{ij}=\frac{\overline{k}_i\overline{k}_j}{\avg{\overline{k}}N}.
\ee
\item 
{\it Structured networks}\\
In the more general structured case we have two possibilities:
\begin{itemize}
\item
{\it i)}
The index $\alpha=1,\ldots,A$ with
$A={\cal O}(N^{1/2})$ can indicate the community of a node and  the function $W(\alpha,\alpha')$ can be a $A\times A$ matrix.
In this case the number of links $L(\alpha,\alpha')$ between the  community $\alpha$ and
the community $\alpha'$  will be distributed
according to a Poisson distribution with average
\be
\overline{L(\alpha,\alpha')}=\sum_{i<j}p_{ij}\delta(\min(\alpha_i,\alpha_j),\alpha)\delta(\max(\alpha_i,\alpha_j),\alpha').
\ee
\item
{\it ii)}
The index $\alpha$ can indicate a position in a metric space which
determine the link probability.
In this case the function $W(\alpha,\alpha')$ is a
 vector depending only on the metric distance $d[\alpha,\alpha']$,
 i.e. $W=W(d[\alpha,\alpha'])$.
\end{itemize}

For structured networks with a generic distribution of $\theta$'s and
a non trivial function of $W(\alpha,\alpha')$ we can consider the
limit when the $[\max_i(\theta)]^2[\max_{\alpha,\alpha'}
W(\alpha,\alpha')]\ll 1$. In this limit the linking probability
$p_{ij}$ given by equation $(\ref{ens})$
reduces to the simple form
\be
p_{ij}=\theta_i\theta_j W(\alpha_i,\alpha_j)
\ee
and we have 
\be
\theta_i=\frac{\overline{k}_i}{\sum_{j\neq i} \theta_j
  W(\alpha_i,\alpha_j)}\simeq \frac{\overline{k}_i}{{\cal N}_{\alpha}}
\ee
with ${\cal N}_{\alpha}=\sum_j \theta_j W(\alpha,\alpha_j)$.

\end{itemize}

\section{Large deviation of  the cluster distribution}

The number of connected components $C({\cal G})$ or "clusters" of a
network ${\cal G}$
gives direct information on the topological structure of the network and
their percolating properties.
Indeed if $C({\cal G})/N$ is small  there are few large connected components
while in the opposite case the network is divided into a huge number of
small clusters.
In the limit of large network sizes $N\rightarrow \infty$ each
canonical generalized network ensemble will be characterized by a
typical value of the number of clusters $C^{\star}(N)$.
The typical distribution of clusters gives the
percolating properties of the  networks belonging to the ensemble and
will be able to characterize the critical exponents of the percolation
phase transition. Moreover different network realizations ${\cal G}$ of a generalized canonical 
ensemble will have a number of
clusters $C({\cal G})$ which is subject  to large deviations with respect to the
typical value $C^{\star}(N)$.

Given the probability of a network $P_c({\cal G})$ in the canonical generalized random
ensembles, as defined in equation $(\ref{canonical})$, we can define 
the probability density $\hat{P}(C)$ of generating a 
random network ${\cal G}$ in this ensemble with $C$ clusters as in the
following:
\begin{equation}\label{probc}
\hat{P}(C)=\sum_{\cal G} P_C({\cal G}) \delta(C,C({\cal G}))\,.
\end{equation}
In the thermodynamic limit, $N\rightarrow\infty$, the probability $\hat{P}(C)$
is centered  at some typical value $C^{\star}$ and decays extremely
fast away from $C^{\star}$ in the large networks limit.
 Let us   indicate with $c=C/N$ the number of connected components per vertex,
the typical value of  this quantity converges in the thermodynamic limit  to a size independent value $c^{\star}$.
Therefore, in order to characterize $\hat{P}(C)$ in the thermodynamic limit, we consider the function $\omega(c)$ defined as
\begin{equation}
 \omega(c)=\lim_{N\to\infty} \frac{1}{N}\ln \hat{P}(C)\,.
\end{equation}
implying clearly  $\omega(c)\leq 0$ for all $c\in [0,1]$
and $\omega(c^*)=0$.\\
Finally we introduce the generating function $Y(q)$ of the cluster probability $\hat{P}(C)$ 
\begin{equation}\label{Yq}
Y(q)=\sum_C \hat{P}(C)q^C=\sum_{{\cal G}({\bf a})}\prod_{ij}
p_{ij}^{a_{ij}}(1-p_{ij})^{1-a_{ij}} q^{C({\cal G})}\,.
\end{equation} 
where in the last expression we have used equation $(\ref{canonical})$ 
defining the generalized random ensembles.
We characterize the asymptotic  limit of the cluster generating 
function $Y(q)$ by the $\phi(q)$ defined as
\begin{equation} \label{phi}
\phi(q)=\lim_{N\to\infty} \frac{1}{N} \ln Y(q) \,.
\end{equation}
From   equation  $(\ref{Yq})$ we obtain, with a saddle point calculation, 
that the conjugated Legendre transform of the quantity $\omega(c)$  can 
be expressed in terms of $\phi(q)$ according to the relation
\begin{equation}\label{leg}
 \omega(c)=\min_q [\phi(q)-c\log q]\,.
\end{equation}
The cluster distribution is therefore fully characterized in the
asymptotic limit if we know the function $\phi(q)$.

\section {The fully connected heterogeneous Potts Model and the
  Percolation transition of the generalized random networks ensembles}

 In this  section we will reduce the problem of finding the
cluster distribution in generalized canonical random ensembles to the study of
a mean-field Potts Models with heterogeneous couplings.
We will prove that $\phi(q)$, given by $(\ref{phi})$,
has a formal relation with the free energy of the mean field Potts
model  with heterogeneous couplings, after a suitable analytic
continuation.
This relation generalizes the known connection between the fully
connected Potts model and  the generating function of the cluster
distribution of a random $G(N,p)$ network \cite{Fortuin,Monasson}.

In order to present the results of the paper in a self-contained way we
describe here the cluster expansion of the fully connected Potts model.
The Potts model is a well known statistical mechanical problem \cite{Potts_rev} describing
$N$ classical degrees of freedom $\sigma_i$ associated to the nodes 
$i=1\ldots N$ of a given network.
Each variable $\sigma_i$ can take $q$ different values, namely 
$\sigma_i=0\ldots q-1$, and is coupled 
to all the other degrees of freedom $\sigma_j$ by means of a two-body 
interaction of strength $J_{ij}$.
This interaction favors configurations where all the nodes in the network 
have the same value of $\sigma$. Thus the energy reads
\begin{equation}
 E[\{\sigma\}] = -\sum_{i<j}\delta(\sigma_i,\sigma_j) J_{ij} -
 h\sum_{\sigma} u_\sigma \sum _i \delta(\sigma,\sigma_i) \,,
\label{E}
\end{equation}
where we assume that all the couplings are positive, $J_{ij}>0$ and
that the first sum in $(\ref{E})$ runs over all the pairs of nodes of
a fully connected network. Moreover, we take the auxiliary
field $h u_\sigma$  parallel to the direction $\sigma$.
The partition function of the model is
\begin{equation}\label{potts}
 Z =\sum_{\{\sigma_i=0,\ldots,q-1\}} \exp\left(-\beta E[\{\sigma\}]\right)\,,
\end{equation}
where $\beta$ is the inverse temperature and the summation runs over 
all $q^N$ spin configurations.
In order to map the Potts model to the cluster structure of the
generalized random network ensembles, we expand the partition
function $Z$ following the article \cite{Fortuin}
\begin{equation}\label{pf}
 Z[\{v_{ij}\},h]=\sum_{\sigma}\prod_{i<j}
 \left[1+v_{ij}\delta(\sigma_i,\sigma_j)\right] e^{\beta h\sum_\sigma u_\sigma \sum_i \delta(\sigma_i,\sigma) }\,.
\end{equation}
where we have defined
\begin{equation}\label{relJv}
v_{ij}=e^{\left(\beta J_{ij}\right)}-1\,.
\end{equation}
Expanding equation (\ref{pf}) we obtain
\begin{eqnarray}\label{cl}
 Z[\{v_{ij}\},h] =  &&\sum_{\sigma}  e^{\beta h\sum_\sigma u_\sigma \sum_i \delta(\sigma_i,\sigma) } \left[ 1+\sum_{ij}
   v_{ij}\delta(\sigma_i,\sigma_j) \right. + \nonumber \\
   &+&\left. \sum_{i<j\;,\;k<l \;(ij)\neq(kl)} 
   v_{ij}v_{kl} \delta
   (\sigma_i,\sigma_j)\delta(\sigma_k,\sigma_l)+\ldots\right]\,.
\end{eqnarray}
Each term in the expansion $(\ref{cl})$ corresponds to a possible network
${\cal G}$ formed by a subset $E({\cal G})$ of edges on the $N$ complete network. Each
contribution from a network ${\cal G}$ is  weighted by the  probability
$\prod_{i j\in E({\cal G})}v_{ij}$ and the sum is made over all possible
networks ${\cal G}$ of $N$ nodes.
Using this expansion,  after performing  the sum over the configurations
$\{\sigma_i=0,\ldots q-1\}$, we can write the partition function reported in 
$(\ref{pf})$, in the form:
\begin{equation}
Z[\{v_{ij}\},h]=\sum_{{\cal G}} \prod_{ij\in E({\cal G})} v_{ij}
\prod_{n=0}^{C({\cal G})-1} \left(\sum_\sigma e^{\beta h u_\sigma
    S_n}\right)\,,
\label{xxx}
\end{equation}
with $E({\cal G})$ given by the set of all edges in ${\cal G}$, $C({\cal G})$
given by the number of connected components in the network and $S_n$ denotes the
size of the $n$-th component.
From the previous equation it follows that in absence of external field
\begin{equation}\label{G}
Z[\{v_{ij}\},h=0]=\sum_{{\cal G}} \prod_{i j\in E({\cal G})} v_{ij} \quad q^{C({\cal
    G})}\,.
\end{equation}
By comparing the definition of the cluster generating function
$(\ref{Yq})$ and the expression $(\ref{G})$ for the  partition function
of the Potts Model, we observe that  the two functions can be related by
the following simple expression:
\begin{equation}\label{equv}
 Y(q)=\prod_{i<j}\left(1-p_{ij}\right) Z[\{v_{ij}=p_{ij}(1-p_{ij})^{-1}\},h=0]\,.
\end{equation}
and the associated logarithmic function reads
\begin{equation}
 \phi(q)= \sum_{i<j} \ln(1-p_{ij}) - f[\{v_{ij}\}]
\end{equation}
where $v_{ij}=p_{ij}(1-p_{ij})^{-1}$ and $f$ is defined at null external field $h=0$.
In the high temperature limit $\beta \rightarrow 0$ the couplings
$J_{ij}$ given by $(\ref{relJv})$ are linked to the edge probability 
by means of the equation (\ref{relJv}) so that
\begin{equation}\label{pijJij}
 v_{ij} = \frac{p_{ij}}{(1-p_{ij})} \sim \beta J_{ij} \,.
\end{equation}
Therefore in order to find the cluster generating function we can simply
solve the fully connected Potts model with heterogeneous couplings. Any assumption on the network ensemble will have a direct counterpart
on the structure of the couplings in the Potts model.\\
We will solve the model in this framework, specializing the results
for the cases of our interest $(\ref{ens})$. Using equation $(\ref{pijJij})$ 
we obtain
\be
v_{ij}=\beta J(\theta_i, \theta_j, \alpha_i, \alpha_j)=\theta_i\theta_j W(\alpha_i,\alpha_j).
\ee
In the various different cases under study the function
$J(\theta_i,\theta_j,\alpha_i,\alpha_j)$ takes different values:
\begin{itemize}
\item
{\it The  $G(N,p)$ ensemble}\\
For the characterization of the cluster distribution of a 
 Poisson random network in the $G(N,p)$ ensemble
 with $p=\avg{k}/N$ we take  
\be
\beta J(\theta_i,\theta_j,\alpha_i,\alpha_j)\simeq \frac{\avg{\overline{k}}}{N} 
\ee 
for all pairs $i,j$.
\item
{\it The  Configuration model}\\
For the characterization of the cluster distribution in the
Configuration model we take  
\be
\beta J(\theta_i,\theta_j,\alpha_i,\alpha_j)= \theta_i \theta_j
\ee
In the case of an uncorrelated network we have $\theta_i \ll 1$ and  
we can express the hidden variables $\theta_i$  in terms of the
expected average degree $\overline{k}_i$, as 
$\theta_i={\overline{k}_i}/{\sqrt{\avg{\overline{k}}N}}$
Consequently the couplings of the Potts model take the form
\be
\beta J(\theta_i,\theta_j,\alpha_i,\alpha_j)=\frac{\overline{k}_i\overline{k}_j}{\avg{\overline{k}}N}.
\ee
\item For the characterization of the cluster distribution 
in structured network ensemble with community structure or spatial
dependence on the embedding geometric space, we have
\be
\beta J(\theta_i,\theta_j,\alpha_i,\alpha_j)=\theta_i \theta_j W(\alpha_i,\alpha_j).
\ee
In the case in which $(\max_i \theta_i)^2[\max_{\alpha,\alpha'}
W(\alpha,\alpha')]\ll 1 $  the previous equation simplifies
\be\label{strcnet}
\beta J(\theta_i,\theta_j,\alpha_i,\alpha_j)=\overline{k}_i
\overline{k}_j \frac{W(\alpha_i,\alpha_j)}{{\cal
    N}_{\alpha_i}{\cal N}_{\alpha_j}}.
\ee
\end{itemize}

For $q \rightarrow 1$ the properties of the partition function (\ref{G})
are in correspondence with the percolation properties \cite{Fortuin} of the
generalized canonical network ensembles with linking probabilities $p_{ij}$
given by (\ref{ens}). We will sketch the proof following \cite{Lub}.
It is straightforward that in the limit $q\to 1$, the partition function
 $Z[\{v_{ij}\},h]=\prod_{i<j}(1-p_{ij})^{-1}$ so that
\be\label{f1}
f_1 = \lim_{N\to \infty \; q\to 1}  \frac{\ln Z[\{v_{ij}\},h]+\sum_{i<j}\ln(1-p_{ij})}{N(q-1)}  =\lim_{N\to \infty} \left.\frac{\partial \ln Z[\{v_{ij}\},h]}{N\partial q}\right|_{q=1}.
\ee
We could choose the parameter $u_\sigma = \delta(\sigma,0)$, so that the
external field favors the $\sigma=0$ state, the partition function 
reported in (\ref{xxx}) simplifies
\be
Z[\{v_{ij}\},h] = \sum_{\cal G} \prod_{ij \in E({\cal G})} v_{ij}\prod_{n=0}^{C({\cal G})-1}\left(q-1+e^{hS_n}\right)\,.
\ee
Using the fact that $\sum_n f(S_n)=\sum_S \sum_n \delta(S_n-S) f(S) =
\sum_s C(S) f(S)$, where $S$ is the number of nodes in the same cluster 
and $C(S)$ the number of clusters with $S$ nodes, we obtain the previous 
equation becomes
\be
Z[\{v_{ij}\},h] = \sum_{\cal G} \prod_{ij \in E({\cal G})} v_{ij}\quad e^{\sum_S C(S) \ln\left(q-1+e^{h S}\right)}\,.
\ee
Performing the summation over the graphs with a 
saddle point approximation, we obtain in the thermodynamic limit the
equation (\ref{f1}) is
\be
f_1= \frac{1}{Z_1}\langle \sum_{s({\cal G})} c(s) e^{-hs}\rangle = \sum_s c(s) e^{-hs}\,,
\ee
where $s=S/N$ and $c(s) =C(S)/N$. Differentiating the previous equation with respect to the external
field we obtain that the node probability to be in the percolating cluster is linked to the free
energy function of the Potts model in the limit $q\to 1$
\be\label{probcluinf}
\lim_{h\to 0^+} 1+\frac{\partial f_1}{\partial h} = 1- \sum_s s c(s) = \mathcal{P}(\{p_{ij}\})\,.
\ee
While the second derivative gives the mean clusters per nodes. Using the Potts model,
we are also able to compute the probability two 
given nodes belong to the percolating component.
Let us introduce the node-node correlation in the limit $h\to 0$
\be\label{correlation}
D_{ij}(q)=\sum_\sigma e^{-\beta E(\{\sigma\})} \delta_{\sigma_i,\sigma_j} = \langle \delta_{\sigma_i\sigma_j}\rangle\,,
\ee
that measures the probability two nodes have the same colour. 
We could easily compute this quantity and we obtain 
\be
\lim_{q\to 1} \frac{q}{q-1} D_{ij}(q) = 1-\langle \mathcal{C}_{ij}\rangle
\ee
where $\mathcal{C}_{ij}$ is the indicator function: if node $i$ and $j$ are in the
same cluster it has the value one, otherwise it vanishes. We want to underline
the fact that the probability two nodes are in the same non-percolating component
is defined through the following relation
\be
\mathcal{H}_{ij} = \langle \mathcal{C}_{ij}\rangle - \mathcal{P}^2(\{p_{kl}\})
\ee
This shows how solving the Potts model in $q\to 1 $ limit, gives
us information on the percolating transition in generalized network ensemble.

\section{Free energy of the Potts model and the percolation phase transition}

In order to solve the mean-field Potts model we introduce the order parameters
\begin{equation}
 c_{\theta\alpha}(\sigma)=\frac{1}{N_{\theta \alpha}}\sum_{i=1}^N \delta(\sigma,\sigma_i) \delta(\theta,\theta_i)\delta(\alpha,\alpha_i)\,.
\end{equation}
where 
\be
N_{\theta\alpha}=\sum_i \delta(\theta,\theta_i)
\delta(\alpha,\alpha_i)
\ee are the number of nodes with a given  \emph{hidden variables} $\theta$ and $\alpha$.
The order parameters $c_{\theta \alpha}$ satisfy  their  proper normalization
\begin{equation}
 \sum_{\sigma} c_{\theta \alpha}(\sigma)=1\,.
\end{equation}
The energy of the Potts model in absence of external
field $h=0$, expressed in terms of the order
parameters $c_{\theta \alpha}(\sigma)$, takes the form
\begin{equation}
E[\{c_{\theta\alpha}(\sigma)\}]=-\frac{N^2}{2} \sum_{\sigma,\theta,\theta',\alpha,\alpha'} p_{\theta\alpha} p_{\theta'\alpha'} c_{\theta\alpha}(\sigma) c_{\theta'\alpha'}(\sigma) J(\theta,\theta',\alpha,\alpha') + O(N)
\end{equation}
where we have explicitly shown the dependence of the coupling from external parameters $\theta$ and $\alpha$.
In order to express the partition function as a sum over the collective variables $c_{\theta\alpha}(\sigma)$,
we need to take into account the entropic contribution, counting the number of microscopic configuration
with a given value of $c_{\theta\alpha}(\sigma)$. To the leading order in $N$ we get
\begin{equation}\label{parfuncc}
 Z = \sum_{\{c_{\theta\alpha}(\sigma)\}} e^{-\beta E[c_{\theta\alpha}(\sigma)]} \prod_{\theta,\alpha}\left( \frac{N_{\theta\alpha}!}{\prod_\sigma [N c_{\theta\alpha}(\sigma)]!}\right)= \sum_{c_{\theta\alpha}(\sigma)} e^{-\beta N f[\{c_{\theta\alpha}(\sigma)\}]}
\end{equation}
where the free energy density functional reads
\begin{eqnarray}\label{freenerg1}
\beta f[\{c_{\theta\alpha}(\sigma)\}] = &-& \frac{N}{2}  \sum_{\sigma \theta \theta' \alpha \alpha'} p_{\theta\alpha} p_{\theta'\alpha'} c_{\theta\alpha}(\sigma) c_{\theta'\alpha'}(\sigma)\beta J(\theta,\theta',\alpha,\alpha')\nonumber\\ &+& \sum_{\sigma \theta \alpha} p_{\theta\alpha} c_{\theta\alpha}(\sigma) \ln c_{\theta\alpha}(\sigma)\,.
\end{eqnarray}
In the large $N$ limit one can evaluate the sum in (\ref{parfuncc}) by
the saddle-point method.
As a function of $q$, the Potts model undergoes a phase transition. For
$q\leq q_c$ the order parameter is invariant under the permutation of
the spin values $\sigma=0,\ldots,q-1$. Nevertheless above the percolation
transition, for $q>q_c$  the ground state breaks the symmetry
of the Hamiltonian.

\subsection{Symmetric saddle point}

The free energy of the Potts model is invariant under the permutation of the $q$ colors.
When this symmetry is also shared by the ground state, the fraction of nodes of a given color could be written as
\begin{equation}\label{sym}
 c_{\theta\alpha}(\sigma)=\frac{1}{q}\,,
\end{equation}
which ensures different colors to be identical. Inserting this ansatz in equation (\ref{freenerg1}) we get
\begin{equation}
\beta f^{sym}(q)= -\frac{N}{2 q} \beta \sum_{\theta \theta' \alpha \alpha'} p_{\theta\alpha}p_{\theta'\alpha'}J(\theta,\theta',\alpha,\alpha')- \ln q\,.
\end{equation}
Computing the second order derivative of the free energy density functional, we can study the stability
of the symmetric solution. When the eigenvalue of the Hessian Matrix
of the free energy $(\ref{freenerg1})$ changes sign and becomes negative 
the ansatz (\ref{sym}) is no more correct. The Hessian matrix reads 
\begin{eqnarray}\label{hessian}
 H_{\theta\theta'\alpha \alpha'}(\sigma, \tau)&&= \frac{\partial^2 \beta
   f[c_{\ell \beta}(\rho)]}{\partial c_{\theta\alpha}(\sigma) \partial
   c_{\theta'\alpha'}(\tau)} = \delta(\sigma,\tau)\times \nonumber  \\ &&
 \left[\delta(\theta,\theta')\delta(\alpha,\alpha')
   \frac{p_{\theta\alpha}}{c_{\theta\alpha}(\sigma)} -
   p_{\theta\alpha}p_{\theta'\alpha'} N \beta J(\theta,\theta',\alpha,\alpha')\right]
\end{eqnarray}
and the related eigenvalue problem is
\begin{equation}\label{eigenvalue}
 (\lambda_{\theta\alpha}-p_{\theta\alpha}q) e_{\theta\alpha} = -p_{\theta\alpha} M_{\theta\alpha}
\end{equation}
where the quantity $M_{\theta\alpha}$ is defined as
\begin{equation}\label{defm}
 M_{\theta\alpha}=\sum_{\theta'\alpha'} p_{\theta'\alpha'} N \beta J(\theta,\theta',\alpha,\alpha') e_{\theta'\alpha'}\,.
\end{equation}
Inserting equation (\ref{eigenvalue}) into (\ref{defm}), we find
\begin{equation}\label{eig}
 M_{\theta\alpha} = -\sum_{\theta'\alpha'} \frac{ p_{\theta'\alpha'}^2
   N \beta J(\theta,\theta',\alpha,\alpha')} {\lambda_{\theta'\alpha'} -p_{\theta'\alpha'} q}M_{\theta'\alpha'}\,,
\end{equation}
defining the eigenvalues of the Hessian matrix in (\ref{hessian}). In order to obtain the critical values for the external parameters that cause instability in the free energy density, we have to find when eigenvalues change sign. Upon imposing $\lambda_{\theta\alpha}=0$ we find this condition is
\begin{equation}\label{matrix}
 M_{\theta\alpha} = \sum_{\theta'\alpha'}
 \frac{1}{q}p_{\theta'\alpha'} N \beta J(\theta,\theta',\alpha,\alpha')M_{\theta'\alpha'}\,.
\end{equation}
In the general case  $\beta J(\theta,\theta',\alpha,\alpha')=\theta \theta'
W(\alpha,\alpha')$, the stability condition can be expressed as
\be
q\leq q_c=\Lambda 
\label{qc}
\ee
with $\Lambda$ indicating the maximal eigenvalue of the matrix
\be
K_{\alpha,\alpha'}={N}\sum_{\theta} p_{\theta\alpha'}  \theta^2 W(\alpha,\alpha').
\label{K}
\ee

In the following we  study in detail the critical point $q_c$ defined by
$(\ref{qc})$ and $(\ref{K})$ in few relevant cases of the generalized
network ensembles.
\begin{itemize}
\item
{\it The $G(N,p)$ ensemble}\\
In the special case of the networks in the $G(N,p)$ ensemble networks with a delta like
distribution $p_\theta=\delta(\theta,\sqrt{\avg{\overline{k}}/{N}})$,
the critical point for percolation $q=1$ provided by the expressions
$(\ref{qc})$ and $(\ref{K})$  is the well known percolation condition for a random network  ${\overline{k}}=\avg{\overline{k}}=1$
\item
{\it The Configuration model}\\ 
In the case of Configuration model the couplings factorize, $\beta J(\theta,\theta',\alpha,\alpha') \sim {\theta \theta'}$.
The stability condition $(\ref{qc})$ $(\ref{K})$ becomes 
\begin{equation}
 q_c=N{\langle \theta^2\rangle}\,.
\label{qcconf}
\end{equation}
In the case in which the network is uncorrelated we have $
\theta_i={\overline{k}_i}/{\sqrt{\avg{\overline{k}}N}}$ and 
 the degree $k_i$ of a node $i$ is a Poisson variable with average
$\overline{k}_i$.
The critical point $(\ref{qc})$ can be then expressed in terms of
the actual degree of the canonical Configuration ensemble as
\be
q_c=\frac{\avg{\overline{k(k-1)}}}{\avg{\overline{k}}}
\ee  
In the typical case limit, i.e. $q=1$, the previous equation corresponds 
to the condition for the percolation transition in Configuration  networks \cite{Cohen1,Noh,Boguna}.

\item
{\it Structured networks}\\
In the general case of the structured networks the complete eigenvalue
problem in equation $(\ref{qc})$  and equation  $(\ref{K})$ have to be solved on a
case by case basis in order 
to find the percolation critical point.

Nevertheless in the following we presents two simple cases in which
the problem can be simplified.
\begin{itemize}
\item {\it First case}\\
We present a case in which a perturbative analysis can give good
approximation to the critical point.
The case under study is the case in which the network has a detailed
structure made of $A$  different communities labeled with an index
$\alpha=1,\ldots, A$ and $A\simeq {\cal O}(1)$. Each community has well
defined features such as the average degree and the number of links 
shared with other communities.
This naturally leads to an interaction between nodes which depends 
on the community they belong to, encoded in the following matrix
\begin{equation}
 W({\alpha, \alpha'})= \left\{ 
\begin{array}{cc}
\psi & \mbox{ if } \alpha=\alpha'\\
\xi & \mbox{ if } \alpha\neq \alpha'
\end{array}\right.
\end{equation}
In this hypothesis the
matrix $K$ $(\ref{K})$ takes the form
\begin{equation}
K_{\alpha \alpha'} = NW(\alpha,\alpha')\sum_\theta p_{\theta\alpha'} \theta^2 =  W({\alpha\alpha'}) \langle \theta^2\rangle_{\alpha'}\,.
\end{equation}
where we indicated with  $\langle x\rangle_\beta=\sum_\theta
p_{\theta \beta} x$ the  average over one single component
$\beta$. 
The eigenvalue problem $(\ref{qc})$ that we have to solve to find the
critical point of the Potts model can be solved perturbatively in the limit $ \Delta ={\xi-\psi} \ll 1$. In this case the matrix $K$ is
\begin{equation}
 K = \psi \left(D + \Delta H\right)
\end{equation}
where $D$ is a diagonal matrix and $H$ has vanishing diagonal elements
\begin{eqnarray}
 D_{\alpha\alpha'}&=& ={N\langle \theta^2\rangle_{\alpha}}\delta(\alpha,\alpha') \nonumber \\
H_{\alpha\alpha'} &=& \left\{
\begin{array}{cc}
0 & \mbox{ if } \alpha=\alpha'\\
N\langle \theta^2\rangle_{\alpha'} & \mbox{ if } \alpha\neq \alpha'
\end{array}\right.
\end{eqnarray}
It is well known in perturbation theory for non degenerate states,
that the eigenvalues of this problem  show second order corrections to
the diagonal entries $D_{\alpha\alpha'}$ in
the parameter $\Delta$. Finally we obtain that the onset of instability 
occurs when  the following relation is satisfied
\begin{eqnarray}
q_c = N\max_{\alpha=1,\ldots,A} {{\psi}} \left[\langle \theta^2\rangle_{\alpha} + \Delta^2 \sum_{\alpha'\neq \alpha} \frac{\langle \theta^2\rangle_{\alpha} \langle \theta^2\rangle_{\alpha'}}{|\langle \theta^2\rangle_{\alpha}-\langle \theta^2\rangle_{\alpha'}|} + o(\Delta^3)\right]\,.
\end{eqnarray}
This set of coupled equations reduce to the value found in the
Configuration model, i.e. $q_c=N\langle \theta^2\rangle$ when there is only one single community. Here we report the condition for the leading term in
$0(\Delta^0)$ that has the following form
\begin{equation}\label{qcconncomp}
q_c= \psi N\max_{\alpha=1,\ldots A}{\langle \theta^2\rangle}_{\alpha}\,.
\end{equation}
We want to underline the new percolation condition becomes
\begin{equation}
N \max_{\alpha=1,\ldots,A}{\langle \theta^2\rangle_\alpha} = \frac{1}{\psi}
\end{equation}
meaning that  the percolation transition depends strongly on the
number of links of the most connected community.\\
Whenever different communities have the same distribution i.e. the same
second moment $\langle \theta^2\rangle_{\alpha} = \langle \theta^2\rangle$, we are able to perform the calculation exactly and the critical value $q_c$ reads 
\be\label{critrel}
q_c = (\psi+(A-1)\phi) N\langle \theta^2\rangle
\ee

\item
{\it Second case}-\\
The second case that we consider is formed by sparse structured
networks  with  the couplings $\beta J(\theta,\theta',\alpha,\alpha')$ taking the expression $(\ref{strcnet})$ that we write here for 
convenience
\be
\beta J(\theta_i,\theta_j,\alpha_i,\alpha_j)=\overline{k}_i \overline{k}_j\frac{W(\alpha_i,\alpha_j)}{{\cal
    N}_{\alpha_i}{\cal N}_{\alpha_j}}
\ee
In the further approximation that the density of nodes with "hidden
variables" $\theta$ and $\alpha$ are factorisable,
i.e. $p_{\theta\alpha}=p_{\theta}\hat{p}_{\alpha}$ we can simplify the
eigenvalue problem  $(\ref{qc})$, $(\ref{K})$ to find the critical
point of the Potts model as 
\be
q_c=\avg{\overline{k(k-1)}} \hat{\Lambda}
\ee 
where $\hat{\Lambda}$ is the maximal eigenvalue of the matrix $\hat{K}$
defined as 
\be
\hat{K}_{\alpha,\alpha'}=N\hat{p}_{\alpha'}
\frac{W(\alpha,\alpha')}{{\cal N}_{\alpha}{\cal N}_{\alpha'}}.
\ee
\end{itemize}
\end{itemize}

\subsection{Asymmetric saddle point}

Below the phase transition  the symmetric solution $(\ref{sym})$ is no 
more stable, as shown in the previous section. 
In the stationary state of the Potts model a giant component appears,
and a more complicated saddle point
has to be found. Due to the fact that one single color becomes dominant, 
generalizing for similar ansatz made for the Potts model with homogeneous
couplings \cite{Monasson},  
 the following ansatz on the parameter $c_{\theta\alpha}$ is proposed
\begin{eqnarray}
c_{\theta\alpha}(\sigma,s_{\theta\alpha})&=&\frac{1}{q}\left(1+(q-1)s_{\theta\alpha}\right) \qquad \mbox{if}\; (\sigma=0)\nonumber\\
c_{\theta\alpha}(\sigma,s_{\theta\alpha})&=&\frac{1}{q}\left(1-s_{\theta\alpha}\right) \qquad \mbox{else }
\end{eqnarray} 
And thus the density functional free energy reads
\begin{eqnarray}\label{free_energy}
 \beta f[\{c_{\theta\alpha}(\sigma,s_{\theta\alpha})\}] &=& \frac{N}{2
  q}\sum_{\theta \theta'\alpha \alpha'} p_{\theta\alpha}
p_{\theta'\alpha'} \beta J(\theta,\theta',\alpha,\alpha') \bigl[(1-q) s_{\theta\alpha} s_{\theta'\alpha'}-1\bigr]+ \nonumber\\ &-&\log q +\sum_{\theta\alpha} \frac{p_{\theta\alpha}}{q} \Bigl[(q-1)(1-s_{\theta\alpha})\log(1-s_{\theta\alpha}) +\nonumber\\ &+& \bigl(1+(q-1) s_{\theta\alpha}\bigr) \log\bigl(1+(q-1)s_{\theta\alpha}\bigr)\Bigr]\,,
\end{eqnarray}
where we have to minimize over the variational parameters $s_{\theta\alpha}$. Solving the equation
\begin{equation}
 \left.\frac{\partial \beta f[\{c_{\theta\alpha}(\sigma,s_{\theta\alpha})\}]}{\partial s_{\theta\alpha}}\right|_{s^*_{\theta\alpha}}=0
\end{equation}
we finally obtain the self consistent condition for the parameter we solved numerically
\begin{equation}\label{selfcons}
 s^*_{\theta\alpha}= \frac{e^{ \theta\rho_{\alpha} }-1}
{q-1+e^{\theta \rho_{\alpha} }}\,.
\end{equation}
with $\rho_{\alpha}$ given by
\be\label{rhodef}
\rho_{\alpha'}=N\sum_{\alpha \theta} {\theta}W(\alpha,\alpha')p_{\alpha\theta} s_{\theta\alpha}. 
\ee 
Therefore equation $(\ref{selfcons})$ can be expressed as a close expression
for $\rho_{\alpha}$, which is the order parameter for the Potts phase
transition. In particular we find
\be
\rho_{\alpha'}=N\sum_{\theta,\alpha} {p_{\theta\alpha} \theta
  W(\alpha,\alpha')} \; \frac{e^{\theta \rho_{\alpha}}-1}{q-1+
  e^{\theta \rho_{\alpha}}}\,.
\label{rho}
\ee
The solution of this equation is $\rho_{\alpha}=0$ for $q\leq q_c$ and
develops a non zero solution  for $q>q_c$.
The transition can be continuous or discontinuous. In all the ensembles
studied in this paper, $q_c= 2$ signs the crossover between 
a second order phase transition and a first order one. This could be
understood in the general framework of Landau Theory.
As it is well known, the Hamiltonian of the Potts model is invariant
under the permutation symmetry of the $q$ colors, which in the case $q \leq 2$ 
 is accidentally equivalent to the $Z_2$ symmetry. 
In equation (\ref{free_energy}) it is easy to show the free
energy is explicitly even under the transformation 
$s_{\theta \alpha}\to -s_{\theta\alpha}$ when $q\leq 2$,
while for higher $q$, the free energy density $f$ contains all possible 
powers of the order parameter $s_{\theta\alpha}$. 
As a consequence, within the Landau Theory, the property of the free energy 
for $q\leq 2 $ necessary reflects into a continuous phase transition 
at least in absence of an external field. 
Thus on general ground we expect the crossover from second to first
order transition could occur only at the value $q=2$ independently 
on the network we choose.
If we expand $(\ref{rho})$ to the first order in $\rho_{\alpha}$
we get the equation
\be
\rho_{\alpha}=\frac{1}{q}\sum_{\alpha'} K_{\alpha,\alpha'}\rho_{\alpha'}
\ee  
with the matrix $K_{\alpha, \alpha'}$ given by $(\ref{K})$.
We recover therefore the same critical point $(\ref{qc})$,
$q_c=\Lambda$ where $\Lambda$ is the maximal eigenvalue of the matrix
$K$ as was found by studying the stability of the Potts model above 
the phase transition.

For the case of the percolation transition, i.e. $q\rightarrow 1$ 
\cite{Fortuin} we have a continuous phase transition and we can study 
equation $(\ref{rho})$ for small values of $\rho_{\alpha}$ to find the 
critical exponents of the percolation transition.
\begin{itemize}
\item
{\it The $G(N,p)$ ensemble}\\
In this case the order parameter $\rho_{\alpha}$ is independent on
$\alpha$ $\rho_{\alpha}=\rho$ and the self-consistent equation $(\ref{rho})$ simplify to
\be
\rho =N  \theta \frac{e^{\theta \rho}-1}{q-1+
  e^{\theta \rho}}
\label{rhognp}
\ee
with $\theta=\sqrt{\avg{\overline{k}}/N}$.
The expansion for small value of $x=\theta\rho$  and $q=1$ gives
\be
x=\avg{\overline{k}}\left({x}-\frac{x^2}{2}\right)
\ee
therefore we can derive the known result that 
\be
\rho\propto (\avg{\overline{k}}-1)^{\hat{\beta}}
\ee
with  the mean field critical exponent  $\hat{\beta}$ given by $\hat{\beta}=1$.
\item
{\it The Configuration model}\\
In the case of the Configuration model
the order parameter $\rho_{\alpha}$ is independent on $\alpha$,
i.e. $\rho_{\alpha}=\rho$ and
the self-consistent equation (\ref{rho}) reduces to 
\begin{equation}
\rho=\sum_\theta \frac{p_\theta \theta}{\langle \theta\rangle} \;
\frac{e^{\theta \rho}-1}{q-1+ e^{\theta \rho}}
\label{rhoc}
\end{equation}
where 
$\rho=\sum_\theta \frac{\theta}{\sqrt{\langle\theta\rangle}}p_\theta s_\theta$. 
The expansion of this equation for small value of $\rho$ provides the
critical exponents for networks in the Configuration model and
generalizes the results of uncorrelated networks to network with the
correlations imposed  by the Configuration model.
In the case in which $\avg{\theta^3}$ is finite, the expansion of
(\ref{rho}) for $q=1$
gives
\bea
\rho&=&N\sum_{\theta}
p_{\theta} \theta^2  \;
\left(\rho-\frac{1}{2}\theta
  \rho^2\right)
\eea
The order parameter close to the percolation phase transition goes
like
\be
\rho\propto (N\avg{\theta^2}-1)^{\hat{\beta}}
\ee
with $\hat{\beta}=1$ as in the $G(N,p)$ ensemble.
On the contrary, in the case  in which  $p_{\theta}\propto\theta^{-\gamma}$ with
$\gamma\in(3,4]$ the expansion of equation $(\ref{rhoc})$ gives
\bea
\rho=N\avg{\theta^2}\rho-\rho^{\gamma-2}I
\eea
 giving the critical exponent $\hat{\beta}=\frac{1}{\gamma-3}$.
Finally we study the case in which $p_{\theta}=\theta^{-\gamma}$ and
$\gamma<3$.
In this case the self-consistent equation $(\ref{rhoc})$ can be
written as
\be
\rho=\rho^{\gamma-2}\hat{I}
\ee
giving $\rho=\hat{I}^{\hat{\beta}}$ with the critical exponent $\hat{\beta}=\frac{1}{3-\gamma}$.
\item
{\it The structured networks}\\
In general the study of the percolation transition for structured
networks might be considered  on case by case basis depending  on the
$p_{\theta \alpha}$ distribution and on the function
$W(\alpha,\alpha')$ under consideration.
Here we consider the case in which the distribution $p_{\theta
  \alpha}$ is factorisable, i.e. $p_{\theta
  \alpha}=p_{\theta}\hat{p}_{\alpha}$.
In this case, if 
the moment  $\avg{\theta^3}$ is finite, expanding equation $(\ref{rho})$ for
$q=1$ we get  
\bea\label{uno}
\rho_{\alpha}&=&N\sum_{\theta,\alpha'}
p_{\theta\alpha} \theta^2 W(\alpha,\alpha') \;
\left(\rho_{\alpha'}-\frac{1}{2}\theta
  \rho_{\alpha'}^2\right)\nonumber \\
&& =\sum_{\alpha'}K_{\alpha,\alpha'}\rho_{\alpha'}-\frac{1}{2}\sum_{\theta\alpha'} p_{\theta\alpha}W(\alpha,\alpha')\theta^3 \rho_{\alpha'}^2
\eea
If we write $\rho_{\alpha}$ in terms of the eigenvectors $\vec{u}^{\lambda}$ of the matrix
$K$, i.e.
\be
\rho_{\alpha}=\sum_{\lambda}c_{\lambda} u^{\lambda}_{\alpha}
\ee
 the equation $(\ref{uno})$ can be written as an equation for the
 constants $c_{\lambda}$. Solving perturbatively assuming that
 $c_{\Lambda}\gg c_{\lambda}$ for $\lambda\neq \Lambda$ with $\Lambda$
given by the maximal eigenvalue of the matrix $K$, we get
\bea
c_{\Lambda}\propto (\Lambda-1) &&\nonumber \\
c_{\lambda}\propto (\Lambda-1)^2 &\mbox{ for }& \lambda \neq \Lambda 
\eea
Therefore in this case there are two  critical exponent $\hat{\beta}=1$ for
the maximal eigenvector and $\hat{\beta}'=2$ for all the other eigenvectors.
In the case $p_{\theta}\propto \theta^{-\gamma}$ with $\gamma\in(3,4]$
we have the expansion
\bea
\rho_{\alpha}&=& =\sum_{\alpha'}K_{\alpha,\alpha'}\rho_{\alpha'}-\sum_{\theta\alpha'} p_{\theta}\hat{p}_{\alpha}W(\alpha,\alpha')H(\alpha) \rho_{\alpha'}^{\gamma-2}
\eea
in this case we have 
\bea
c_{\Lambda}\propto (\Lambda-1)^{1/(\gamma-3)} && \nonumber \\
c_{\lambda}\propto (\Lambda-1)^{2/(\gamma-3)}&\mbox{ for }&\lambda\neq \Lambda
\eea
and the critical exponent $\hat{\beta}=1/(\gamma-3)$ and
$\hat{\beta}'=2/(\gamma-3)$ which generalizes the
results for scale-free networks.
In the case in which $p_{\theta}\propto \theta^{-\gamma}$ and $\gamma<3$
the expansion of the equation $(\ref{rhoc})$ gives
\bea
\rho_{\alpha}=\sum_{\alpha'}\rho_{\alpha'}^{\gamma-2}W(\alpha,\alpha')\hat{H}_{\alpha'}.
\eea
getting the critical exponents $\hat{\beta}=1/(3-\gamma)$ and 
$\hat{\beta}'=(4-\gamma)/(3-\gamma)$.
\end{itemize} 
Using equation ($\ref{selfcons}$) and $(\ref{rhodef})$ we find an 
explicit expression of the free energy of the Potts model
in the asymmetric phase (\ref{free_energy}) as a function of the 
order parameter vector $\rho_{\alpha}$
\begin{eqnarray}\label{fqasg}
\beta f(q) = &-&\frac{1}{2q}\sum_{\theta\theta'\alpha\alpha'} p_{\theta\alpha}p_{\theta'\alpha'}N\theta \theta' W(\alpha,\alpha') - 
\frac{q-1}{q} \sum_{\theta\;\alpha} p_{\theta\alpha} \theta \rho_\alpha 
\frac{e^{\theta\rho_{\alpha}-1}}{q-1+e^{\theta\rho_{\alpha}}}\nonumber\\
 &+&\sum_{\theta\;\alpha}p_{\theta\alpha} \left[\frac{\theta 
\rho_\alpha e^{\theta\rho_\alpha} }{q-1+e^{\theta\rho_\alpha}} - 
\log\left(q-1+e^{\theta\rho_\alpha}\right)\right]
\end{eqnarray}
Using this expression we find for the function $\phi(q)$ the explicit
expression as a function of $\rho_{\alpha}$, 
\begin{eqnarray}\label{phiqasg}
&\phi(q)&=\frac{1-q}{2q}\left[\sum_{\theta\theta'\alpha\alpha'} p_{\theta\alpha}p_{\theta'\alpha'}N\theta \theta' W(\alpha,\alpha') - \sum_{\theta\;\alpha} p_{\theta\alpha} \theta \rho_\alpha 
\frac{e^{\theta\rho_{\alpha}-1}}{q-1+e^{\theta\rho_{\alpha}}}\right]\\\nonumber
 &-&\sum_{\theta\;\alpha}p_{\theta\alpha} \left[\frac{\theta 
\rho_\alpha e^{\theta\rho_\alpha} }{q-1+e^{\theta\rho_\alpha}} - 
\log\left(q-1+e^{\theta\rho_\alpha}\right)\right]
\end{eqnarray}
In the specific case of the Configuration model the precedent
equations $(\ref{fqasg})$ and $(\ref{phiqasg})$ can be simplified to the following equation for the free
energy density $f(q)$
\begin{equation}
\beta f(q)= \frac{\langle \theta\rangle}{2 q}\left( (1-q) \rho^2-1\right) + \sum_\theta p_\theta \left[ \frac{\theta\rho e^{\theta\rho} }{q-1+e^{\theta\rho}} - \log(q-1+e^{\theta\rho})\right]
\end{equation}
and the following equation for the function $\phi(q)$.
\begin{equation}
 \phi(q)= \frac{\langle \theta \rangle (q-1)}{2 q}\left(\rho^2-1\right)-\sum_\theta p_\theta \left[\frac{\theta\rho e^{\theta\rho} }{q-1+e^{\theta\rho}}- \log( q-1+e^{\theta\rho} )\right].
\end{equation}

\section{Cluster distribution}

We large deviations $\omega(c)$ of the clusters distribution can be
calculated using $(\ref{leg})$, by performing a Legendre transformation
of the function $\phi(q)$ calculated by evaluating expression $(\ref{phiqasg})$
at the self-consistent solution of equation $(\ref{rho})$,
\begin{equation}
 \omega(c)= \min_q\left( \phi(q) -c \log q\right).
\end{equation}
Therefore we have shown that by solving the heterogeneous Potts model
with couplings $\beta J(\theta,\theta',\alpha,\alpha')=\theta \theta'
W(\alpha,\alpha')$ we can directly characterize the critical point of
the percolation phase transition, the critical exponent
$\hat{\beta}$ of this transition and the large deviation function of the
number of clusters $c=C/N$ present in the networks of the ensemble.
\begin{figure}[ht]
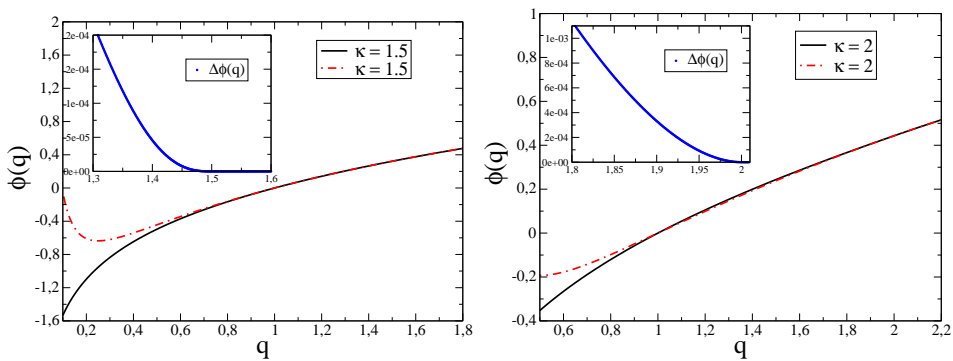

\begin{center}
 \epsfig{figure=Figure/figure1.eps ,scale=0.23}
 \epsfig{figure=Figure/figure2.eps,scale=0.23}
\caption{ The two branches  of the $\phi(q)$ function
 across the Potts model phase transition for
the hidden variables ensembles with average degree distributed
according to a Poissonian. The solid line indicate the function
$\phi(q)$ calculated at  the asymmetric solution of the Potts 
model and the dot-dashed line
indicate the function $\phi(q)$ calculated for the symmetric
solution. For $\kappa \leq 2$ the free energy at the phase transition
varies continuously, from the  $\rho=0$ solution to the asymmetric
solution. The inset show the difference of the free energy calculated
on the two branches. 
In the left figures $\kappa=1.5$ while in the right $\kappa=2$. }
\label{figure1}
\end{center}
\end{figure}

\begin{figure}[ht]
\begin{center}
\epsfig{figure=Figure/figure3.eps ,scale=0.23}
\epsfig{figure=Figure/figure4.eps,scale=0.23}
\caption{The two  branches of the $\phi(q)$ function across the Potts model
 phase transition for the Configuration model with the average of the
degrees at each node distributed according to a Poissonian.  The solid
line indicate the function $\phi(q)$  calculated at  the asymmetric 
solution of the Potts model and the dot-dashed line
indicate the function $\phi(q)$ calculated for the symmetric
solution. The free energy at the transition varies discontinuously at
$q_c\ge \kappa=\avg{\overline{k}}+1$ when the metastable Configuration
$\rho\neq0$ disappears.  The inset shows explicitly the discontinuity
in the difference
$\Delta \phi$ between the  function $\phi$ calculated on the two
solutions.}
\label{figure2}
\end{center}
\end{figure}

\begin{figure}[ht]
\begin{center}
 \epsfig{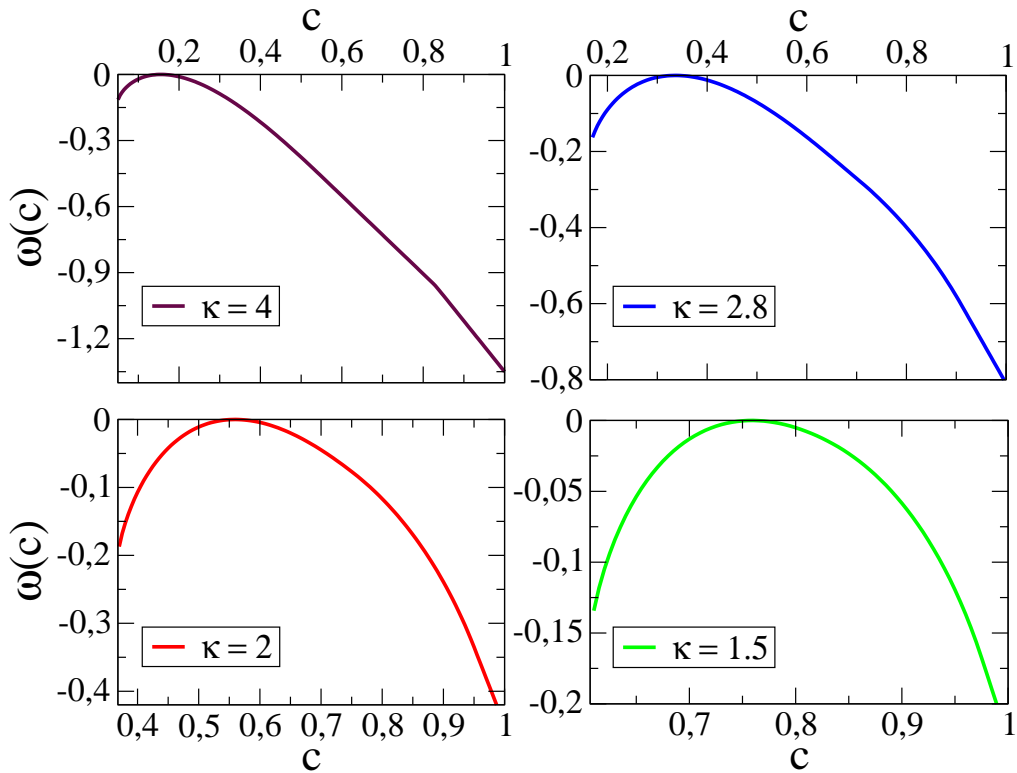} 
\caption{We show the numerically evaluated $\omega(c)$ function for
the hidden variable Configuration model in which the average degrees
are distributed according to a Poisson distribution with
$\kappa=1+\avg{\overline{k}}$. We show the dependence on the typical number
$c^*$ as a function of the mean connectivity. The higher is  $\kappa$
the smaller the number of expected typical components.}
\label{omega}
\end{center}
\end{figure}

\section{Numerical results}
In this section we present the study of the large deviation of cluster
distribution for different examples of generalized network ensembles.
 \begin{itemize}
\item
{\it The $G(N,p)$  ensemble}\\
We consider the simple case of a $G(N,p)$ networks ensembles where the
average degree of each node $\theta_i$ is independent on $i$, i.e. $
\theta_i =\sqrt{\avg{\overline{k}}/N}$, i.e.
$p=\avg{\overline{k}}/N$. The equation $(\ref{rhognp})$ is the
self-consistent equation for the Potts
model phase transition.
This equation has been already studied  \cite{Monasson} where it was  found that the Potts model has phase transition in $q_c=\avg{\overline{k}}$ of second order for value of $q_C\leq 2$ and of first order for $q_c>2$. 
We suggest the reader to refer to references \cite{Fortuin,Monasson} for a full
account of this case.

\item
{\it The Configuration model}\\
In particular we study the limit of weak heterogeneity when we assume
that the average degree of the nodes $\{\overline{k}\}$ is Poisson
distributed and the case of strong heterogeneity of the degree of the
nodes when the hidden variables $\{\theta_i\}$ are distributed
according to a power-law.

\begin{itemize}
\item
{\it Poisson hidden variable distribution}\\
We consider  the case in which the distribution of the mean values
for the connectivity's $\overline{k}_i$ is Poisson and
$\theta_i=\sqrt{\overline{k}_i/N}$. This ensemble introduce a   small
 heterogeneity with respect to pure Erd\"os and R\'enyi networks.   
In this case the critical point is equal to  $q_c=N\langle
\theta^2\rangle =\avg{\overline{k}}+1$. Therefore  the percolation
transition happens at $\avg{\overline{k}}+1=1$
 revealing that the percolating
phase is   already when
the mean connectivity is $\avg{\overline{k}} \rightarrow
0$.

As predicted by the theoretical results  we found a phase transition
for $q_c=N\avg{\theta^2}=\avg{\overline{k}}+1$ 
depending  on the mean value of connectivity $\langle
\overline{k}\rangle$. In figures \ref{figure1} and \ref{figure2}
we show the function $\phi(q)$ as a function of $q$ for different
values of the parameter $\kappa=\avg{\overline{k}}+1$ that has the same
role of the inverse temperature in the associated Potts model.
 The phase transition is of the second order when $q_c=\langle
\overline{k}\rangle +1 \leq 2$ (See figure \ref{figure1}). On the contrary, 
when  $\langle \overline{k}\rangle +1 > 2$, (See figure \ref{figure2}).
\begin{figure}
\begin{center}
\epsfig{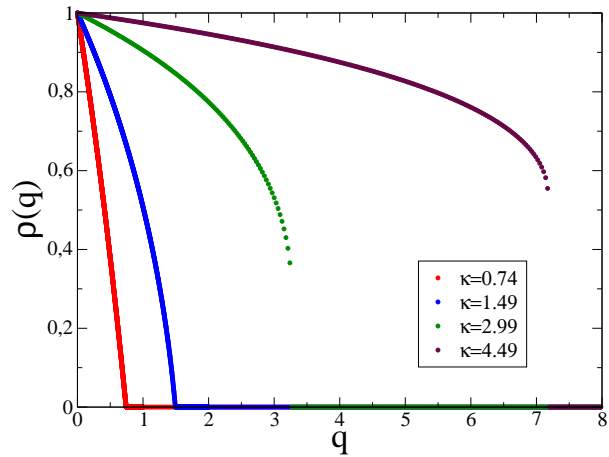}
\caption{The solution of the self-consistent equation $(\ref{rhoc})$ for the
parameter $\rho$  as a function of the number of colors $q$ for a
scale-free  random networks with different values of the critical
point $q_c=\kappa=\avg{\overline{k(k-1)}}{\avg{\overline{k}}}$
and power-law exponent $\gamma=5$. }
\label{alpha}
\end{center}
\end{figure}

\begin{figure}[ht]
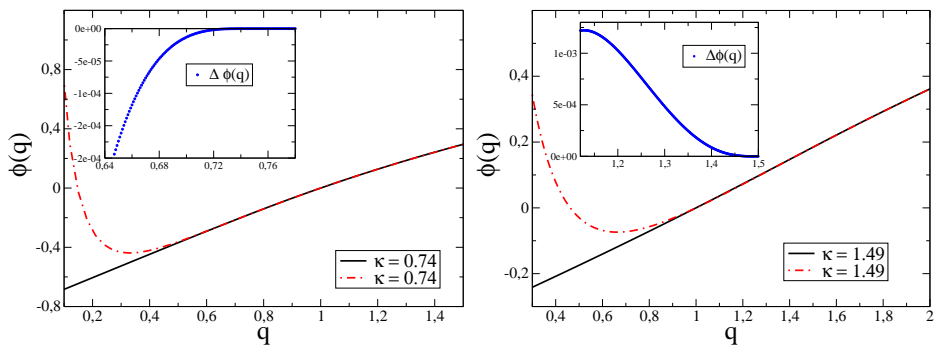

\begin{center}
\epsfig{file=Figure/figure7.eps,scale=0.23}
\epsfig{file=Figure/figure8.eps,scale=0.23}
\caption{The two branches of the $\phi(q)$ function for a 
Configuration model with in power-law distribute $\{\theta\}$'s with 
exponent $\gamma = 5$. The solid
line is the function $\phi(q)$  calculated at  the asymmetric 
solution of the Potts model and the dot-dashed line
 is the function $\phi(q)$ calculated for the symmetric
solution. The transition point is at 
$q_c=\kappa=N\langle \theta^2\rangle\leq 2$. In the inset we show the
difference in free energy in the two branches which is continuous.}
\label{scalefree1}
\end{center}
\end{figure}
In figure $(\ref{omega})$ we show the function $\omega(c)$ for this ensemble
for different values of  $\kappa=1+\avg{\overline{k}}$.We obtain that for value
of $c\leq \hat{c}=e^{-\avg{\overline{k}}}$ the function
$\omega(c)$ tends to infinity, i.e $\omega(c) \to-\infty$. This value is relative to the minimum
number of components that are equivalent to the number of isolated
vertices. Therefore we plot the function $\omega(c)$ only for $c>\hat{c}$.
As a function  of the parameter $\langle \overline{k}\rangle$ we found
that different number of connected components are dominant and in
particular for $\langle \overline{k}\rangle +1> 2$ we find that the typical
number of cluster are lesser that 0.5 and that in the limit of high 
$\avg{\overline{k}}$ this number vanishes (See figure \ref{omega}). 

\item {\it Scale-free degree distribution}\\
 We use as degree distribution $p_\theta\sim \theta^{-\gamma}$ with
 $\gamma>3$ so that the second moment does not vanish. We fixed the 
exponents $\gamma$, letting the value of infrared cutoff changes in
order to fix all the moments $\langle \theta^m\rangle$.
In figure \ref{alpha} we show the behavior of the solution $\rho$ of
the self-consistent equation $(\ref{rhoc})$ as
a function of the parameter $q$ for scale-free networks.
When $q<q_c$ defined by equation $(\ref{qc})$, i.e. 
$q_c=\avg{\overline{k(k-1)}}/\avg{\overline{k}}$ a non zero solution
$\rho\neq 0$ is found while for $q>q_c$ the $\rho=0$ solution becomes the
stable one. The value of the stable solution $\rho$ of equation
$(\ref{rhoc})$ close to the phase transition $q\simeq q_c$ varies
continuously in the case $q_c\leq 2$ and discontinuously for $q>2$.
There is a second order phase transition for values of $\frac{\langle
\overline{k(k-1)}\rangle}{\langle \overline{k}\rangle}\leq 2$
and a first order phase transition for $\frac{\langle
\overline{k(k-1)}\rangle}{\langle \overline{k}\rangle}> 2$
where the free energy discontinuously goes to zero as shown in figure
\ref{scalefree1}.

In figure \ref{scalefree1} and \ref{scale2}  we plot the difference of
the $\phi(q)$ functions calculated on the two branches of the
solution of $(\ref{rhoc})$ (the solution with  $\rho=0$ and the other
non trivial solution stable for $q<q_c$). From these figures we can see that $\Delta
\phi(q)$ has a  discontinuity in the regime $\frac{\langle
\overline{k(k-1)}\rangle}{\langle \overline{k}\rangle}>2 $ and
vanishes continuously in the regime  $\frac{\langle \overline{k(k-1)}\rangle}
{\langle \overline{k}\rangle}\leq 2 $. 
\begin{figure}[ht]
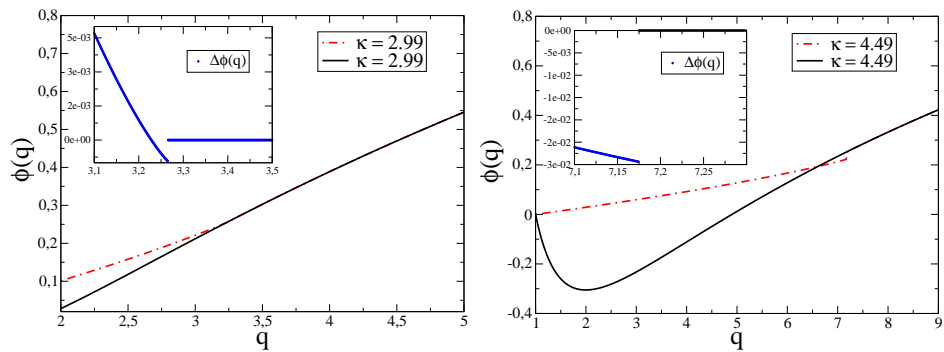

\begin{center}
\epsfig{file=Figure/figure9.eps,scale=0.23}
\epsfig{file=Figure/figure10.eps,scale=0.23}
\caption{The two branches of the  $\rho(q)$ function for value of
parameter $\kappa=N\langle \theta^2\rangle >2$ for 
scale free network $\gamma = 5$. 
In the inset  we show the difference between the free energy associated 
to the symmetric solution  $\rho=0$ and the free energy associated to the other
asymmetric solution with $\rho \neq 0$. In the figure we report the
difference of the free energy calculated on the two branch solution showing
evidence for  the
discontinuity  in the free energy at the transition point.}
\label{scale2}
\end{center}
\end{figure}
In figure \ref{omegascale} is shown the probability of large deviation
in the number of clusters for the Configuration model with power-law
distributed value of $\{\theta\}'s$. The typical number of clusters
$c^{\star}$ is a decreasing function of
$\kappa=\avg{\overline{k(k-1)}}/\avg{\overline{k}}$.
\begin{figure}[ht]
\begin{center}
 \epsfig{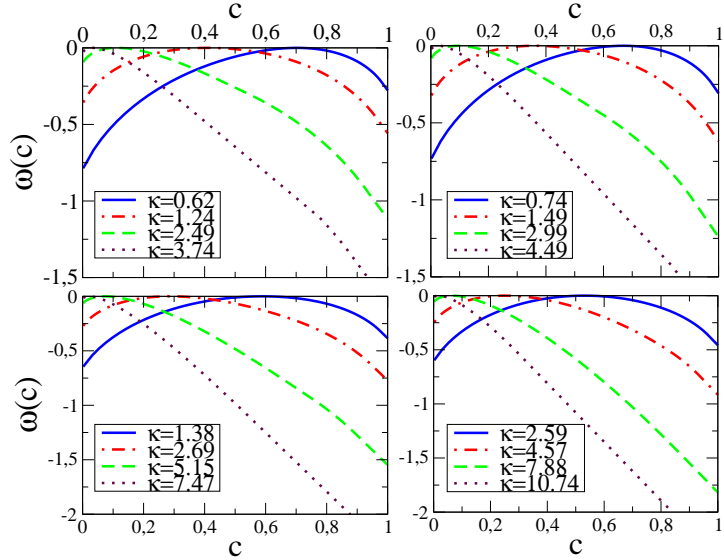}
\caption{The logarithm of the probability density function $\omega(c)$ 
versus the number of connected components $c$. It is easy to identify the typical value $c^{\star}$ when the function $\omega$ touches the zero axis. 
This number  depends strongly on $\kappa=N\langle \theta^2\rangle$. 
Each network correspond to different choice of the parameter 
$\gamma=7,5, 3.5,3.01$ starting from the top on the left.}
\label{omegascale}
\end{center}
\end{figure}
\begin{figure}
\begin{center}
\epsfig{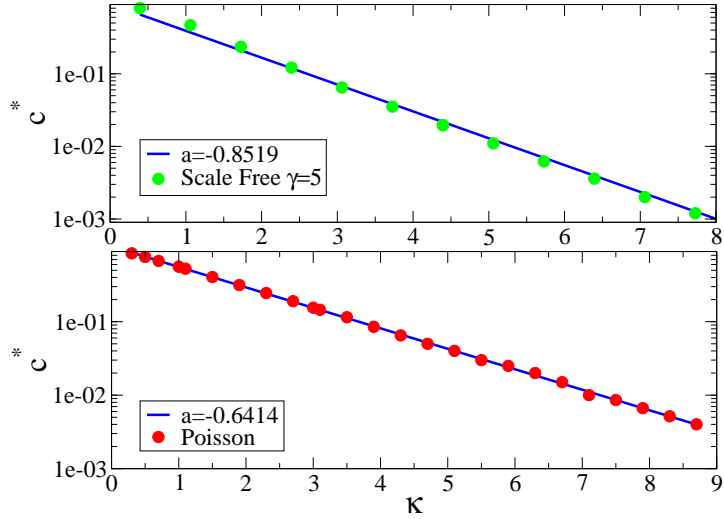}
\caption{The typical number of clusters  per vertex $c^{\star}$ vanishes
exponentially fast with the  increasing  of $\kappa=N\langle
 \theta^2\rangle$. Here we show in a logarithmic scale this relation
for Poissonian degree distribution and power-law networks for value a
value of  $\gamma=5$. We report also the value from the best fitting 
calculation that has a good agreement with the data points.}
\label{cluster}
\end{center}
\end{figure}
\item {\it Comparison of the typical number of cluster for Configuration
model  with Poisson and power-law distributed $\{\theta\}$'s.}\\
The typical value of the number of clusters $c^{\star}$ depends on the parameter $N\langle \theta^2\rangle$. This dependencies is
shown in figure \ref{cluster}. In the case of power-law distributed
$\{\theta\}'s$, the characteristic scale above which the number $c^*$
vanishes is higher than in Poisson case. 

\end{itemize}

\item {\it Simple case of structure network: Network with four equal communities}

\begin{figure}[ht]
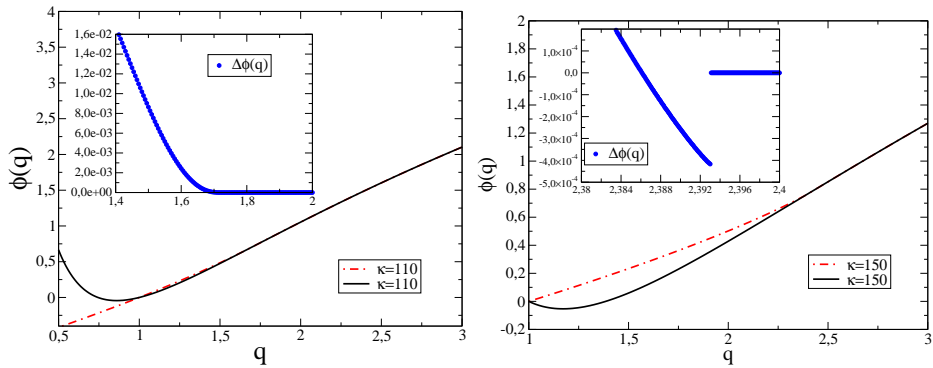

\begin{center}
\epsfig{file=Figure/figure13.eps,scale=0.23}
\epsfig{file=Figure/figure14.eps,scale=0.23}
\caption{ In figure we plot $\phi(q)$ on the two branches of the solution for
network with four communities (second case under study). The network
ensemble is characterized by a  
  $W(\alpha, \alpha')$ given by (\ref{interaction}) with $x=0.5$ and
  different values of $\theta$,  $\theta=110$ and $\theta= 150$. This network show the same crossover between a first order 
and a second order phase transition as soon as the parameter
$\theta>\theta_c$ calculated in equation $\ref{zzz}$, $\theta_c=120$.The solid
line indicate the function $\phi(q)$  calculated at  the asymmetric 
solution of the Potts model and the dot-dashed line
indicate the function $\phi(q)$ calculated for the symmetric
solution. 
 This is more clear by the discontinuity in the difference of $\phi(q)$ calculated in 
two branches of the solution reported on the inset. The solid line 
corresponds to $\rho=0$ while the dashed one is $\phi(q)$ calculated on
the non trivial solution $\rho\neq 0$.}
\label{freecc}
\end{center}
\end{figure}

\begin{figure}[ht]
\begin{center}
\epsfig{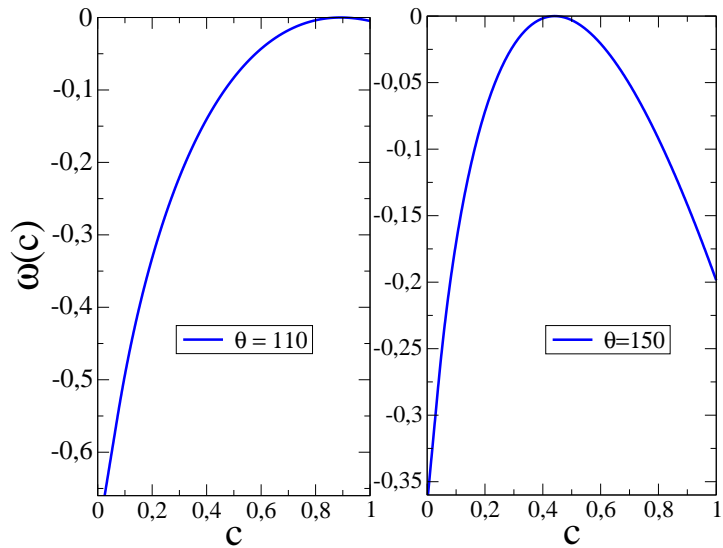}
\caption{We report the logarithm of the cluster probability
  $\omega(c)$ for the study network ensemble with four communities
  (second case under study). The network ensemble is characterized by a 
  $W(\alpha, \alpha')$ given by (\ref{interaction}) with $x=0.5$ and
  the two different values of $\theta$, $\theta=110$ and $\theta=150$. The distribution 
is dominated by the typical number $c^*$ that depends strongly on the average
connectivity of the network ensembles.}
\label{omegacc}
\end{center}
\end{figure}

In general the case of structure networks with non trivial
$W(\alpha,\alpha')$ functions have to be studied on a case by case
approach.
\begin{itemize}
\item{\it First case-}
Here we consider the particular example in which the network is
divided in four equivalent communities.
This networks have been considered as a benchmark networks with
community structure \cite{Newman}.
We consider in particular the case with $N \theta_i=\mbox{const}\quad \forall i$ and
the network divided in four communities $\alpha=1,2,3,4$, i.e. $A=4$
with 
\bea\label{interaction}
W(\alpha,\alpha')=\left\{\begin{array}{lcr}\ \frac{x}{64}&\mbox{for }&
    \alpha=\alpha' \\ \frac{(1-x)}{192} &\mbox{for}
    &\alpha\neq \alpha'\end{array}\right.
\eea 
In this case we find that for $q > 1/64$ the only stable solution is the zero
one while for lower value of the parameter $q$ there is a non zero solution
that is stable everywhere independently on the value of $x$. 
We want to emphasize the fact that, substituting 
equation (\ref{interaction}) in (\ref{critrel}), we obtain the critical relation reads
\be
q_c = \frac {1}{64}\,,
\ee
showing there is no phase transition for every value of the parameter $x$ at
fixed $\theta$ and $x$.
\item{\it Second case-}
We consider the case with four communities $\alpha=1,2,3,4$, 
i.e. $A=4$ with $W(\alpha, \alpha')$ given by $(\ref{interaction})$ at
fixed $x$ for the value of  the \emph{ hidden variables} 
$\theta_i = \theta \quad \forall i \in A $. The behaviour in that case 
is exactly the same as in the Configuration model. There is a crossover between first and second
order phase transition governed by the parameter $\theta$. The threshold 
value is given by the relation
\be
\theta_c=2(\psi+(A-1)\phi)^{-1}
\label{zzz}
\ee 
and substituting the value of the parameter $x=0.5$ and $A=4$, the critical value is of order $\theta_c\sim 120$. The free energy is shown in figure \ref{freecc} where
it is easily to catch the nature of the phase transition. For completeness we report 
also the logarithm of the clusters distribution in
figure \ref{omegacc} that show the same behaviour of Configuration model.
\end{itemize}
\end{itemize}

\section{Conclusions}
In conclusion, we have studied the percolation transition and the large deviation of the cluster
distribution in generalized canonical random network ensembles. The calculation has been
performed by mapping the problem of finding the cluster distribution on a fully connected
Potts model with heterogeneous couplings. The results generalize the known results for
uncorrelated configuration models to correlated configuration models and are able to predict
the behavior of the phase transition for generic structured networks with non-trivial community
or spatial structure. Ongoing work will investigate what is the role of the percolation properties
in generalized random network ensembles for the understanding of the critical behavior of
dynamical models defined on them and for the characterization of their small loops distribution.

\section{Acknowledgments}
This paper was supported by the project IST STREP GENNETEC contract No. 034952 and by MIUR grant 2007JHLPEZ.\\
\bibliographystyle{unsrt}
\bibliography{biblio}

\begin{thebibliography}{10}

\bibitem{Dorogovtsev}
S.N. Dorogovtsev, A.~V. Goltsev, and J.~F.~F. Mendes.
\newblock {\em Rev. Mod. Phys.}, 80:1275, 2008.


\bibitem{MR}
M. Molloy and B. A. Reed. \newblock{\em Rand. Struc. Algor.} 6:161, 1995.

\bibitem{Attack}
R.~Albert, H.~Jeong, and A.-L. Barab\'asi.
\newblock {\em Nature}, 406:378, 2000.

\bibitem{Cohen1}
R.~Cohen, K.~Erez, D.~ben Avraham, and S.~Havlin.
\newblock {\em Phys. Rev. Lett.}, 85:4626, 2000.

\bibitem{Cohen2}
R.~Cohen, K.~Erez, D.~ben Avraham, and S.~Havlin.
\newblock {\em Phys. Rev. Lett.}, 86:3682, 2001.

\bibitem{Ising1}
S.N. Dorogovtsev, A.~V. Goltsev, and J.~F.~F. Mendes.
\newblock {\em Phys. Rev. E}, 66:016104, 2002.

\bibitem{Ising2}
M.~Leone, A.~Vazquez, A.~Vespignani, and R.~Zecchina.
\newblock {\em Eur. Phys. J. B}, 28:191, 2002.

\bibitem{Ising3}
G.~Bianconi.
\newblock {\em Phys. Lett. A}, 303:166, 2002.

\bibitem{Isaac}
N.~S. Skantzos, I.~Pérez Castillo, and J.~P.~L. Hatchett.
\newblock {\em Phys. Rev. E}, 72:066127, 2005.

\bibitem{Coolen}
A.~C. C.~Coolen et~al.
\newblock {\em J. Phys. A: Math. and Gen.}, 38:8289, 2005.

\bibitem{Noh}
J.~D. Noh.
\newblock {\em Eur. Phys. J. B}, 66:251, 2008.

\bibitem{Bollobas}
B.~B. Bollobas.
\newblock {\em Random graphs}.
\newblock Cambridge, University Press, 2001, 2nd ed. edition, 1985.

\bibitem{Monasson}
Engel, Monasson, and Hartmann.
\newblock {\em J. Stat. Phys.}, 117(3):387, 2004.

\bibitem{Fortuin}
C.~M. Fortuin and P.W. Kasteleyn.
\newblock {\em Physica}, 57:536, 1972.

\bibitem{Boguna}
M.~Boguna and M.~A. Serrano.
\newblock {\em Phys. Rev. E}, 72:016106, 2005.

\bibitem{Doro2}
A.~V. Goltsev, S.~N. Dorogovtsev, and J.~F.~F. Mendes.
\newblock {\em Phys. Rev. E}, 78:051105, 2008.

\bibitem{entropy1}
G.~Bianconi.
\newblock {\em Europhys. Lett.}, 81:28005, 2008.

\bibitem{entropy2}
G.~Bianconi.
\newblock {\em http://arXiv:0802.2888}, 2008.

\bibitem{hv1} 
J. Park and M. E. J. Newman. \newblock{\em Phys. Rev. E} { 70}:
066146, 2004.
\bibitem{hv2} 
B. Sodeberg. 
\newblock{\em Phys. Rev. E} { 66}:066121, 2002.
\bibitem{hv3} 
F. Chung and L. Lu.\newblock{\em  PNAS} { 100}:6313, 2002.
\bibitem{hv4} 
G. Caldarelli, A. Capocci, P. De Los Rios and
M. A. Mu\~noz.
\newblock{\em Phys. Rev. Lett.} { 85}:5468, 2002.
\bibitem{hv5} 
M. Bogu\~n\'a and R. Pastor-Satorras. \newblock{Phys. Rev. E}
{ 68}:036112, 2003.



\bibitem{Lee}
D.~S. Lee, K.~I. Goh, B.~Kahng, and D.~Kim.
\newblock {\em Nuclear Physics B}, 696(3):351 -- 380, 2004.

\bibitem{Potts_rev}
F.~Wu.
\newblock {\em Rev. Mod. Phys.}, 54:235, 1982.

\bibitem{Lub}
T.~C. Lubensky.
\newblock {\em Thermal and geometrical critical phenomena in random systems}, in {\em La Mati\`ere Mal Condens\'ee Les Houches 1978} ed.  R Balian, R Maynard and G Toulouse  
\newblock New York : North-Holland Pub. Co., 1979.


\bibitem{Newman}
M.~Girvan and M.~E.~J. Newman.
\newblock {\em Proc. Natl. Acad. Sci.}, 99:7821, 2002.

\end{thebibliography}

\end{document}